\documentstyle[12pt]{article}
\newcommand{\ret}{\nonumber \\}
\newcommand{\Section}[1]%
{\section{#1}\setcounter{equation}{0}%
\setcounter{theorem}{0}}
\newtheorem{theorem}{Theorem}

{\par\noindent{\em #1:\ }}%
{~\rule{2mm}{2mm}\par\bigskip}
\begin{document}
%%%%%%%%%%%%%%%%%%%%%%%%%%%%%%%%%%%%%%%%%%%
\newpage\thispagestyle{empty}
{\topskip 2cm
\begin{center}
{\Large\bf Friedel Sum Rule as a Trace Formula\\} 
\bigskip\bigskip\bigskip\bigskip
{\large Mahito Kohmoto}\\
\bigskip
{\small \it The Institute for Solid State Physics, The University of Tokyo, 
5-1-5 Kashiwanoha, Kashiwa, Chiba 277-8581, JAPAN}\\
\bigskip
{\large Tohru Koma}\\
\bigskip
{\small \it Department of Physics, Gakushuin University, 
Mejiro, Toshima-ku, Tokyo 171-8588, JAPAN}\\
\smallskip
{\small\tt e-mail: tohru.koma@gakushuin.ac.jp}
\end{center}
%This is also for double spacing
%\newpage
\vfil
\noindent
We examine the Friedel sum rule which states 
that the ``excess charge" due to a single impurity potential 
in a metal is equal to a sum of phase shifts 
for scatterings of electrons by the impurity.
For finite volume, the ``excess charge" is given by the difference between total 
numbers of levels in the Fermi sea with and without the impurity potential. 
However, a sequence of the ``excess charge" for 
finite volume is not necessarily bounded in 
the infinite volume limit, as was pointed out by Kirsch. 
In order to circumvent this difficulty, we define ``excess charge" 
directly for the infinite volume.
The Friedel sum rule is proven to hold  for the ``excess charge" thus defined.

\par\noindent
\bigskip
\hfill
\vfil}\newpage
%%%%%%%%%%%%%%%%%%%%%%%%%%%%%%%%%%%%%%%%%%%
%%%%%%%%%%%%%%%%%%%%%%%%%%%%%%%%%%%%%%%%%%%
%\tableofcontents
%\newpage
%%%%%%%%%%%%%%%%%%%%%%%%%%%%%%%%%%%%%%%%%%%
\Section{Introduction}

In this paper, we examine the Friedel sum rule \cite{Faulkner,Friedel,Kittel}. 
Consider a metal with a single impurity at zero temperature. 
Electrons are scattered by the impurity potential, and 
the charge distribution of electrons below the Fermi level 
changes. ``Excess charge" is defined to be 
the difference between the total numbers of levels in the Fermi sea  
with and without the impurity potential for a fixed Fermi energy $E_{\rm F}$.  
Then the Friedel sum rule states that the ``excess charge" 
can be expressed in terms of a sum of phase shifts due 
to the impurity potential. However, a sequence of the ``excess charge" for 
finite volume is not bounded in 
the infinite volume limit, as was pointed out by Kirsch \cite{Kirsch}. 
Although this result may not necessarily imply that there exists no sequence 
of finite boxes such that the sequence of the corresponding ``excess charges" 
converges to a finite value in the infinite volume limit, 
it clearly implies that the ``excess charge" cannot be well-defined 
in the infinite-volume limit from a sequence of finite boxes. 
In order to circumvent this difficulty, we define ``excess charge" directly 
in the infinite volume. We prove that 
the ``excess charge" thus defined can be expressed in terms of 
a sum of phase shifts of scatterings of electrons by 
the impurity potential. Thus we justify the Friedel sum rule in the infinite volume. 

In the next section, we describe the model and our results.   
The proofs of the main theorems are given in Sections~\ref{proof:Theorem1} 
and \ref{proof:Theorem2}. 

%%%%%%%%%%%%%%%%%%%%%%%%%%%%%%%%%%%%%%%
\Section{Model and the results}
\label{model}

Consider a single electron in an impurity potential $V$ 
on ${\bf R}^3$. The Hamiltonian $H$ is 
\begin{equation}
H=-\Delta +V.
\label{H}
\end{equation} 
We assume that $V$ is bounded and of compact support. 
Let $P$ be the projection on energies smaller than the Fermi energy $E_{\rm F}$. 
We denote by $H_0=-\Delta$ the Hamiltonian without the impurity potential, 
and denote by $Q$ the corresponding projection on energies smaller than 
the same Fermi energy $E_{\rm F}$. We also denote by $x=(x_1,x_2,x_3)\in{\bf R}^3$ 
the coordinate of the electron. 
Let $\chi_R$ be the characteristic function of the ball defined by 
$B_R=\{x\in{\bf R}^3|\>|x|\le R\}$ centered at $x=0$ with the radius $R>0$, 
where $|x|:=\sqrt{x_1^2+x_2^2+x_3^2}$. We define the excess charge by 
\begin{equation}
Z:=\lim_{R\uparrow\infty}{\rm Tr}\>\chi_R(P-Q)\chi_R.
\label{Z}
\end{equation}

\begin{theorem} 
\label{theorem1}
The limit, $\lim_{R\uparrow\infty}{\rm Tr}\>\chi_R(P-Q)\chi_R$, exists 
for all the Fermi energy $E_{\rm F}>0$. 
\end{theorem}

The proof will be given in Section~\ref{proof:Theorem1}. 
\smallskip

The Friedel sum rule \cite{Friedel,Kittel} gives a relation between 
excess charge $Z$ and the number of bound states and 
phase shifts for a spherical potential $V$.  
The precise statement is as follows:  

\begin{theorem}
\label{theorem2}
Assume that the impurity potential $V$ is spherical, i.e., 
$V$ is a function of the single variable $|x|$. 
Let $N_b(E_{\rm F})$ be the number of the bound states of the Hamiltonian $H$ with 
the energy smaller than the Fermi energy $E_{\rm F}>0$, 
and let $\eta_\ell(k)$ be the phase shifts of the scattering states 
with angular momentum $\ell$ and with wavenumber $k$. 
We choose the phase shift $\eta_\ell(k)$ to be continuous in all $k$ 
with $\lim_{k\rightarrow \infty}\eta_\ell(k)=0$ for all $\ell$. 
Then 
\begin{equation}
\lim_{R\uparrow\infty}{\rm Tr}\>\chi_R(P-Q)\chi_R
=N_b(E_{\rm F})+\frac{1}{\pi}\sum_{\ell=0}^\infty(2\ell+1)
\left[\eta_\ell(k_{\rm F})-\eta_\ell(0)\right],
\label{FSR}
\end{equation}
where $k_F$ is the Fermi wavenumber, i.e., $k_{\rm F}=\sqrt{E_{\rm F}}$.  
\end{theorem}

The proof will be given in Section~\ref{proof:Theorem2}. 
\smallskip

\noindent
{\bf Remark:} 1. Since the Friedel formula (\ref{FSR}) states that 
the difference between the two bulk quantities is related to   
the boundary quantities, the formula can be geometrically interpreted 
as an analogue of the Gauss or Stokes theorem .  
\smallskip

\noindent
2. In one and two dimensions, one can easily obtain a similar 
formula to the Friedel sum rule (\ref{FSR}) for the phase shifts. 
However, our method cannot be applied to the systems in higher dimensions $d>3$ 
because of the singularity of the Green function at short distance. 
See eq.~(\ref{singG0}) and the corresponding argument 
in the next Section~\ref{proof:Theorem1} for details.  
\smallskip

\noindent
3. For the present spherical potential $V$, one can prove the absence of bound states 
with the binding energy $E=0$ and with the angular momentum $\ell=0$. 
Further, the Levinson theorem \cite{Levinson} implies 
that\footnote{See, for example, Chap.~12 of 
the book \cite{Newton} of Newton.} 
\begin{equation}
\eta_0(0)=0\quad\mbox{for }\ \ell=0,\quad
\mbox{and}\quad(2\ell+1)\eta_\ell(0)=\pi N_\ell\quad\mbox{for }\ 
\ell\ge 1,
\end{equation}
where $N_\ell$ is the number of bound states with angular momentum $\ell\ge 1$, 
including those with zero binding energy. Substituting these into 
the above result (\ref{FSR}), we have the well-known form of 
the Friedel sum rule \cite{Friedel,Kittel} as 
\begin{equation}
\lim_{R\uparrow\infty}{\rm Tr}\>\chi_R(P-Q)\chi_R
=\frac{1}{\pi}\sum_{\ell=0}^\infty(2\ell+1)\eta_\ell(k_{\rm F})
\label{FSRD}
\end{equation}
for the Fermi energy $E_{\rm F}>0$. 
\smallskip

\noindent
4. We should remark the relation between the excess charge $Z$ 
and the ``Krein spectral shift". 
Actually the excess charge $Z$ is equal to 
the spectral shift in a formal calculation. 
For the spectral shift, see recent papers of Combes-Hislop-Nakamura \cite{CHN} 
and Hundertmark-Simon \cite{HS} and references therein.

%%%%%%%%%%%%%%%%%%%%%%%%%%%%%%%%%%%%%%%%%%%%%%%%
\Section{Proof of Theorem~\ref{theorem1}}
\label{proof:Theorem1}

We begin with recalling the well known fact that the spectrum $\sigma(H)$ of 
the present Hamiltonian $H$ of (\ref{H}) is absolutely continuous 
in the region of the positive energy. 
See, for example, the book \cite{ReedSimonIII}. 
The projections below the Fermi level $E_{\rm F}>0$ are, respectively, written  
\begin{equation}
P=\frac{1}{2\pi i}\int_\Gamma \frac{dz}{z-H}\quad\mbox{and}\quad
Q=\frac{1}{2\pi i}\int_\Gamma \frac{dz}{z-H_0} 
\end{equation}
by using the contour integral and the resolvent. 
{From} this expression, one has 
\begin{equation}
P-Q=\frac{1}{2\pi i}\int_\Gamma\>dz\left(\frac{1}{z-H}-\frac{1}{z-H_0}\right). 
\end{equation}
The difference of the two resolvents in the integrand 
is computed as 
\begin{eqnarray}
\frac{1}{z-H}-\frac{1}{z-H_0}&=&\frac{1}{z-H_0}V\frac{1}{z-H}\ret
&=&\frac{1}{z-H_0}V\frac{1}{z-H_0}+\frac{1}{z-H_0}V\frac{1}{z-H}V
\frac{1}{z-H_0}.
\end{eqnarray}
These observations yield  
\begin{eqnarray}
{\rm Tr}\>\chi_R(P-Q)\chi_R&=&\frac{1}{2\pi i}\int_\Gamma\>dz\>
{\rm Tr}\>\chi_R\frac{1}{z-H_0}V\frac{1}{z-H_0}\chi_R\ret
&+&\frac{1}{2\pi i}\int_\Gamma\>dz\>{\rm Tr}\>\chi_R\frac{1}{z-H_0}V\frac{1}{z-H}V
\frac{1}{z-H_0}\chi_R.
\label{TrP-Qresolv}
\end{eqnarray}
It is not difficult to evaluate this right-hand side \cite{Simon} except for 
the contributions of the contour integral near the Fermi level. 
In the following, we treat only the contributions near the Fermi level. 
Therefore we write $z=(k_{\rm F}+i\mu)^2$ near the Fermi level $E_{\rm F}$ 
by introducing the real variable $\mu$. 
Then the integral kernel of the resolvent $(z-H_0)^{-1}$ with $z=(k_{\rm F}+i\mu)^2$ 
is written 
\begin{equation}
G_0(x;(k_{\rm F}+i\mu)^2)=-\frac{1}{4\pi|x|}\exp[(ik_{\rm F}-\mu)|x|],
\label{Green0}
\end{equation}
and the contribution of the contour integral near the Fermi level is written as 
\begin{equation}
\frac{1}{2\pi i}\int dz\cdots=
\lim_{\varepsilon\downarrow 0}\frac{1}{\pi}\int_{-a}^{-\varepsilon}
+\int_{+\varepsilon}^{+a} d\mu\times(k_{\rm F}+i\mu)\cdots 
\end{equation}
with a positive cutoff $a$.  

Using the representation (\ref{Green0}) of the Green function, we have  
\begin{eqnarray}
& &{\rm Tr}\>\chi_R\frac{1}{z-H_0}V\frac{1}{z-H_0}\chi_R\ret
&=&\frac{1}{16\pi^2}\int d^3x\int d^3x'\>\chi_R(x)\frac{1}{|x-x'|^2}
\exp\left[2(ik_{\rm F}-\mu)|x-x'|\right]V(x').
\label{TrRvR}
\end{eqnarray}
In passing, we remark the following: 
In a higher dimension $d>3$, the Green function $G_0(x;z)$ behaves as 
\begin{equation}
G_0(x;z)\sim\frac{{\rm Const.}}{|x|^{d-2}}
\label{singG0}
\end{equation}
for the short distance $|x|\sim 0$. In that case, the right-hand side 
of (\ref{TrRvR}) becomes a divergent integral. Therefore our method 
cannot be applied to the higher dimensional systems.  

We return to the present three dimensional system. 
In this case, the contribution for $x$ in finite range in the integral 
of the right-hand side of (\ref{TrRvR}) gives a finite value.  
Therefore it is sufficient to evaluate the following integral: 
\begin{equation}
I_1=\int_\varepsilon^a(k_{\rm F}+i\mu)d\mu 
\int_{B_R\backslash B_{R'}}d^3 x
\frac{1}{|x-x'|^2}
\exp\left[2(ik_{\rm F}-\mu)|x-x'|\right],
\label{I1}
\end{equation}
where we have chosen the radius $R'$ of the ball $B_{R'}$ to satisfy 
$B_{R'}\subset B_R$ and $B_{R'/2}\supset {\rm supp}\>V$, and 
we have taken account of the contour integral near the Fermi level 
with the cutoff constants $a,\varepsilon>0$ as mentioned above. We will take 
the limit $\varepsilon\downarrow 0$. Note that
\begin{eqnarray}
\int_\varepsilon^a(k_{\rm F}+i\mu)d\mu \exp[-2\mu|x-x'|]
&=&\left.\left[-\frac{1}{2|x-x'|}\exp[-2\mu|x-x'|](k_{\rm F}+i\mu)
\right]\right|_\varepsilon^a\ret
&+&\frac{i}{2|x-x'|}\int_\varepsilon^a d\mu \exp[-2\mu|x-x'|]\ret
&=&\frac{k_{\rm F}+i\varepsilon}{|x-x'|}\exp[-2\varepsilon|x-x'|]+{\cal O}(|x-x'|^{-2})\ret
\label{integralbypart} 
\end{eqnarray}
for a large $|x-x'|$. Substituting this into (\ref{I1}), we have 
\begin{eqnarray}
\lim_{\varepsilon\downarrow 0}I_1&=&
\lim_{\varepsilon\downarrow 0}\int_{B_R\backslash B_{R'}}d^3 x
\frac{k_{\rm F}+i\varepsilon}{|x-x'|^3}\exp\left[2(ik_{\rm F}-\varepsilon)|x-x'|\right]
+{\cal O}(1)\ret
&=&\int_{B_R\backslash B_{R'}}d^3 x
\frac{k_{\rm F}}{|x-x'|^3}\exp\left[2ik_{\rm F}|x-x'|\right]
+{\cal O}(1).
\label{I1expand} 
\end{eqnarray}
By taking the radial variable $r=|x-x'|$ centered at $x'$, 
the integral about the radius $r$ in the first term can be written  
\begin{equation}
\int_{R_0}^\infty dr \frac{1}{r} e^{2ik_{\rm F}r}
\end{equation}
in the limit $R\uparrow\infty$, where $R_0$ is a cutoff constant given by 
$R_0=n_0\pi k_{\rm F}^{-1}$ with a positive integer $n_0$. 
Since both of the real and imaginary parts of the integral can be treated in 
the same way, we consider only the imaginary part. 
This integral converges to a finite value as  
\begin{eqnarray}
\int_{R_0}^\infty dr \frac{\sin 2k_{\rm F}r}{r}
&=&\sum_{n=2n_0}^\infty\int_{r_n}^{r_{n+1}}dr\frac{\sin 2k_{\rm F}r}{r}\ret
&=&\sum_{m=n_0}^\infty
\left[\int_{r_{2m}}^{r_{2m+1}}dr\frac{\sin 2k_{\rm F}r}{r}
+\int_{r_{2m+1}}^{r_{2m+2}}dr\frac{\sin 2k_{\rm F}r}{r}\right]\ret
&=&\sum_{m=n_0}^\infty
\int_{r_{2m}}^{r_{2m+1}}dr\>\sin 2k_{\rm F}r 
\left(\frac{1}{r}-\frac{1}{r+\pi k_{\rm F}^{-1}/2}\right)\ret
&=&\sum_{m=n_0}^\infty
\int_{r_{2m}}^{r_{2m+1}}dr\>\sin 2k_{\rm F}r 
\frac{\pi k_{\rm F}^{-1}/2}{r(r+\pi k_{\rm F}^{-1}/2)}<\infty,\ret
\end{eqnarray}
where $r_n=n\pi k_{\rm F}^{-1}/2$ with the integer $n\ge n_0$. 

The integrand of the second term in the right-hand side of (\ref{TrP-Qresolv}) 
is written   
\begin{eqnarray}
& &{\rm Tr}\>\chi_R\frac{1}{z-H_0}V\frac{1}{z-H}V
\frac{1}{z-H_0}\chi_R\ret
&=&\frac{1}{16\pi^2}\int d^3x \int d^3x' \int d^3 x'' 
\chi_R(x)\frac{1}{|x-x'|}\exp[(ik_{\rm F}-\mu)|x-x'|]\ret
&\times&V(x')G(x',x'';(k_{\rm F}+i\mu)^2)V(x'')
\frac{1}{|x''-x|}\exp[(ik_{\rm F}-\mu)|x''-x|],
\label{integral2}
\end{eqnarray}
where $G(x,x';(k_{\rm F}+i\mu)^2)$ is the integral kernel of 
the resolvent $(z-H)^{-1}$. Since the contribution of the integral for $x$ in 
finite range is finite, we consider only 
the integral for $x$ in the region $B_R\backslash B_{R'}$. 
We choose $R'$ large enough so that 
the ball $B_{R'/2}$ includes the support of the potential $V$. Then we have 
\begin{eqnarray}
& &\int_{B_R\backslash B_{R'}}d^3x \int d^3x' \int d^3 x'' 
\frac{1}{|x-x'|}\exp[(ik_{\rm F}-\mu)|x-x'|]\ret
&\times&V(x')G(x',x'';(k_{\rm F}+i\mu)^2)V(x'')
\frac{1}{|x''-x|}\exp[(ik_{\rm F}-\mu)|x''-x|]\ret
&=&\int_{B_R\backslash B_{R'}}d^3x\frac{1}{|x|^2}
e^{2(ik_{\rm F}-\mu)|x|}\left\langle u,V(z-H)^{-1}Vv\right\rangle\ret
&+&\int_{B_R\backslash B_{R'}}d^3x\frac{1}{|x|^3}
e^{2(ik_{\rm F}-\mu)|x|}\left\langle {\tilde u}(x,\cdot),V(z-H)^{-1}V
{\tilde v}(x,\cdot)\right\rangle
\label{integral2d}
\end{eqnarray}
by using  
\begin{equation}
\frac{1}{|x-x'|}=\frac{1}{|x|}+{\cal O}(|x|^{-2})
\end{equation}
and 
\begin{equation}
|x-x'|=|x|-\frac{x\cdot x'}{|x|}+{\cal O}(|x|^{-1})
\end{equation}
for a large $|x|$. Here $z=(k_{\rm F}+i\mu)^2$, and the functions $u$ and $v$ are 
given by  
\begin{equation}
u(x')=\exp[(ik_{\rm F}+\mu)x\cdot x'/|x|]
\end{equation}
and 
\begin{equation}
v(x'')=\exp[(-ik_{\rm F}+\mu)x\cdot x''/|x|],
\end{equation}
and ${\tilde u}(x,x')$ and ${\tilde v}(x,x')$ are bounded functions. 
Note that $\Vert\chi_A(z-H)^{-1}\chi_A\Vert$ with $z=(k_{\rm F}+i\mu)^2$ is bounded 
uniformly \cite{Branges} in $\mu>0$ for the Fermi energy $E_{\rm F}>0$   
and for a compact subset $A$ of ${\bf R}^3$, where $\chi_A$ is the characteristic 
function of $A$. From this and the method in (\ref{integralbypart}), 
the contribution form the second integral in the right-hand side of (\ref{integral2d}) 
can be handled in the same way. 
Thus it is sufficient to evaluate the integral, 
\begin{equation}
I_2=\int_\varepsilon^a(k_{\rm F}+i\mu)d\mu
\int_{B_R\backslash B_{R'}} d^3x\frac{1}{|x|^2}\exp[2(ik_{\rm F}-\mu)|x|]
\left\langle u,V(z-H)^{-1}Vv\right\rangle,
\label{I2}
\end{equation}
with $z=(k_{\rm F}+i\mu)^2$. Integrating by parts, we have 
\begin{eqnarray}
I_2&=&-(k_{\rm F}+i\mu)
\int_{B_R\backslash B_{R'}} d^3x\frac{1}{2|x|^3}\exp[2(ik_{\rm F}-\mu)|x|]
\left.\left\langle u,V(z-H)^{-1}Vv\right\rangle\right|_\varepsilon^a\ret
&+&\int_\varepsilon^ad\mu \int_{B_R\backslash B_{R'}} d^3x
\frac{1}{2|x|^3}\exp[2(ik_{\rm F}-\mu)|x|]
\frac{\partial}{\partial\mu}(k_{\rm F}+i\mu)\left\langle u,V(z-H)^{-1}Vv\right\rangle.\ret
\label{I20}
\end{eqnarray}
The first term is written  
\begin{equation}
-(k_{\rm F}+i\varepsilon)\int_{B_R\backslash B_{R'}} d^3x\frac{1}{|x|^3}
\exp[2(ik_{\rm F}-\varepsilon)|x|]
\left\langle u,V(z-H)^{-1}Vv\right\rangle+{\cal O}(1)
\label{I21}
\end{equation}
with $z=(k_{\rm F}+i\varepsilon)^2$. 
Note that $\left\langle u,V(z-H)^{-1}Vv\right\rangle$ does not depend on 
the radius $|x|$ 
and is bounded uniformly in $\varepsilon>0$ 
for the Fermi energy $E_{\rm F}>0$. 
{From} these observations and the same argument as that 
for the integral of (\ref{I1expand}), 
we have that the first term in (\ref{I21}) 
converges to a finite constant 
as $R\uparrow\infty$ and $\varepsilon\downarrow 0$. 

Similarly, the second term in the right-hand side of (\ref{I20}) is written 
\begin{equation}
-\int_\varepsilon^ad\mu (k_{\rm F}+i\mu)^2\int_{B_R\backslash B_{R'}} d^3x
\frac{1}{2|x|^3}\exp[2(ik_{\rm F}-\mu)|x|]
\left\langle u,V(z-H)^{-2}Vv\right\rangle+{\cal O}(1). 
\end{equation}
Therefore it is sufficient to show that 
$|\left\langle u,V(z-H)^{-2}Vv\right\rangle|$ is bounded uniformly in $\mu>0$. 
Using the resolvent identity, one has  
\begin{equation}
V\frac{1}{z-H}\frac{1}{z-H}V=
V\frac{1}{z-H_0}\frac{1}{z-H}V+V\frac{1}{z-H}V\frac{1}{z-H_0}\frac{1}{z-H}V.
\end{equation}
Moreover, one obtains 
\begin{equation}
V\frac{1}{z-H_0}\frac{1}{z-H}V=V\frac{1}{z-H_0}\frac{1}{z-H_0}V
+V\frac{1}{z-H_0}\frac{1}{z-H_0}V\frac{1}{z-H}V. 
\end{equation}
{From} these two identities, it is sufficient to show boundedness 
of the operator,  
\begin{equation}
V\frac{1}{z-H_0}\frac{1}{z-H_0}V. 
\end{equation}
Using the spectral decomposition, one has 
\begin{eqnarray}
\left\langle \varphi, (z-H_0)^{-2}\psi\right\rangle
&=&\frac{1}{(2\pi)^3}\int_0^\infty d\rho\int_{S^2}d\omega
\frac{1}{[\rho^2-(k_{\rm F}+i\mu)^2]^2}  
f(\rho,\omega)\rho^2\ret
&=&\frac{1}{(2\pi)^3}\int_{k_{\rm F}-\delta}^{k_{\rm F}+\delta}d\rho\int_{S^2}d\omega
\frac{1}{[\rho^2-(k_{\rm F}+i\mu)^2]^2}  
f(\rho,\omega)\rho^2+{\cal O}(1)\ret 
\end{eqnarray}
for two functions, $\varphi,\psi$, of compact support, where 
we have introduced the polar coordinate $k=\rho\omega$ with $\rho=|k|$, 
$\delta$ is a positive cutoff, and 
the function $f$ is given by the product of the Fourier transforms 
of $\varphi,\psi$.  
Note that 
\begin{equation}
\frac{1}{[\rho^2-(k_{\rm F}+i\mu)^2]^2}=-\frac{1}{2\rho}
\frac{\partial}{\partial\rho}\frac{1}{\rho^2-(k_{\rm F}+i\mu)^2}.
\end{equation}
Using this identity and integration by parts, one has 
\begin{equation}
\int_{k_{\rm F}-\delta}^{k_{\rm F}+\delta}d\rho 
\frac{1}{[\rho^2-(k_{\rm F}+i\mu)^2]^2}  
f(\rho,\omega)\rho^2=\frac{1}{2}\int_{k_{\rm F}-\delta}^{k_{\rm F}+\delta}d\rho
\frac{1}{\rho^2-(k_{\rm F}+i\mu)^2}
\frac{\partial}{\partial\rho}f(\rho,\omega)\rho+{\cal O}(1).
\label{rhoint}
\end{equation}
The integral of the first term in the right-hand side is written 
\begin{equation}
\int_{k_{\rm F}-\delta}^{k_{\rm F}+\delta}d\rho
\frac{1}{\rho^2-(k_{\rm F}+i\mu)^2}{\tilde f}(k_{\rm F},\omega)
+\int_{k_{\rm F}-\delta}^{k_{\rm F}+\delta}d\rho
\frac{1}{\rho^2-(k_{\rm F}+i\mu)^2}
\left[{\tilde f}(\rho,\omega)-{\tilde f}(k_{\rm F},\omega)\right],
\end{equation}
where we have written
\begin{equation}
{\tilde f}(\rho,\omega)=\frac{\partial}{\partial \rho}f(\rho,\omega)\rho.
\end{equation}
Note that 
\begin{equation}
\frac{1}{\rho^2-(k_{\rm F}+i\mu)^2}
=\frac{1}{2(k_{\rm F}+i\mu)}
\left[\frac{1}{\rho-k_{\rm F}-i\mu}-\frac{1}{\rho+k_{\rm F}+i\mu}\right]
\end{equation}
and  
\begin{equation}
\frac{1}{\rho-k_{\rm F}-i\mu}=\frac{\rho-k_{\rm F}}{(\rho-k_{\rm F})^2+\mu^2}
+\frac{i\mu}{(\rho-k_{\rm F})^2+\mu^2}. 
\end{equation}
Combining this with the differentiability of the function $f$, 
we can prove that the right-hand side of (\ref{rhoint}) is bounded uniformly 
in $\mu>0$. 
Thus the double limit $R\uparrow\infty$ and $\varepsilon\downarrow 0$ 
for the integral (\ref{I2}) exists.   
Consequently the limit, $\lim_{R\uparrow\infty}{\rm Tr}\>\chi_R(P-Q)\chi_R$, 
exists for the Fermi energy $E_{\rm F}>0$.

%%%%%%%%%%%%%%%%%%%%%%%%%%%%%%%%%%%%%%%%%%%%%%%%
\Section{Proof of Theorem~\ref{theorem2}}
\label{proof:Theorem2}

We give the proof, following the idea of Friedel \cite{Friedel}. 
We denote by $J:=-ix\times\nabla$ the angular momentum, and define 
a regularized excess charge as 
\begin{equation} 
Z_{\beta,R}:={\rm Tr}\>e^{-\beta J^2}\chi_R(P-Q)\chi_R\quad\mbox{for }\> \beta>0.
\label{ZbetaR} 
\end{equation}
Since the integral kernel of $e^{-\beta J^2}$ which is the heat kernel on 
the two-sphere $S^2$ depends only on 
the angular variables except for the identity about the radial part, 
the existence of $\lim_{R\rightarrow\infty}Z_{\beta,R}$ is justified in 
the same way as in Section~\ref{proof:Theorem1}, and 
we recover the excess charge $Z$ of (\ref{Z}) in the limit $\beta\downarrow 0$ as 
$Z=\lim_{\beta\downarrow 0}\lim_{R\rightarrow\infty}Z_{\beta,R}$. 

The integral kernel of the projection $P$ for the Hamiltonian 
$H$ with the impurity potential $V$ is written 
\begin{equation}
P(x,x')=\sum_{n:E_n<E_{\rm F}}\>v_n(x)v_n^{\>\ast}(x')
+\frac{1}{\pi}\int_0^{k_{\rm F}}dk\sum_{\ell,m}u_{k,\ell,m}(x)
u_{k,\ell,m}^{\quad\ast}(x')
\end{equation}
in terms of the bound state $v_n$ and the scattering state $u_{k,\ell,m}$ 
with the angular momentum $\ell$ and the magnetic quantum number $m$. 
Similarly, 
\begin{equation}
Q(x,x')=\frac{1}{\pi}\int_0^{k_{\rm F}}dk\sum_{\ell,m}u_{k,\ell,m}^{\quad (0)}(x)
{u_{k,\ell,m}^{\quad (0)}}^\ast(x') 
\end{equation}
in terms of the scattering state $u_{k,\ell,m}^{\quad (0)}$ of 
the Hamiltonian $H_0$.  
The scattering state $u_{k,\ell,m}$ is written 
\begin{equation}
u_{k,\ell,m}(x)=\frac{f_{k,\ell}(r)}{r}Y_\ell^m(\theta,\phi)
\label{ukellm}
\end{equation}
in terms of the angular part $Y_\ell^m$ and the radial part $f_{k,\ell}$. 
The radial part follows from the equation, 
\begin{equation}
\frac{d^2}{dr^2}f_{k,\ell}+
\left[k^2-V(r)-\frac{\ell(\ell+1)}{r^2}\right]f_{k,\ell}=0.
\end{equation}
Multiplying this by another solution $f_{k',\ell}$ with a different 
wavenumber $k'$, and then interchanging the variables $k$ and $k'$ in 
the resulting equation, and finally taking the difference between 
the two equations, one has 
\begin{equation}
\frac{d}{dr}\left[f_{k',\ell}\frac{d}{dr}f_{k,\ell}-
f_{k,\ell}\frac{d}{dr}f_{k',\ell}\right]
+(k^2-{k'}^2)f_{k',\ell}f_{k,\ell}=0.
\end{equation}   
Integrating over $r$ from $0$ to $R$, one obtains 
\begin{equation}
f_{k',\ell}(R)\frac{d}{dr}f_{k,\ell}(R)-
f_{k,\ell}(R)\frac{d}{dr}f_{k',\ell}(R)=({k'}^2-k^2)\int_0^R dr 
f_{k',\ell}(r)f_{k,\ell}(r), 
\end{equation}
where we have used the fact $f_{k,\ell}(0)=0$. Further, by dividing 
by $k'-k$ and taking the limit $k'\rightarrow k$ in the both sides, one has 
\begin{equation}
I_{k,\ell}^R:=\frac{1}{2k}
\left[\frac{d}{dk}f_{k,\ell}(R)\cdot\frac{d}{dr}f_{k,\ell}(R)-
f_{k,\ell}(R)\frac{d}{dk}\frac{d}{dr}f_{k,\ell}(R)\right]
=\int_0^R dr\left|f_{k,\ell}(r)\right|^2.
\label{IkellR} 
\end{equation}
We will justify differentiability of the function $f_{k,\ell}$ 
with respect to $k$ later.  
The function $f_{k,\ell}(r)$  for a large $r$ outside the support of $V(r)=0$ 
is written as 
\begin{equation}
f_{k,\ell}(r)=\sqrt{2}\left[krj_\ell(kr)\cos\eta_\ell(k)-krn_\ell(kr)
\sin\eta_\ell(k)\right]
\label{exprfkell}
\end{equation}
in terms of the Bessel functions $j_\ell$ and $n_\ell$. 
The asymptotic forms of the Bessel functions for $\ell\ge 1$ are given by 
\begin{equation}
j_\ell(\rho)=\frac{1}{\rho}
\left[\sin(\rho-\pi\ell/2)+\frac{(\ell+1)!}{(\ell-1)!}\frac{1}{2\rho}
\cos(\rho-\pi\ell/2)\right]+{\cal O}(\rho^{-3})
\label{jasymp}
\end{equation}
and
\begin{equation}
n_\ell(\rho)=-\frac{1}{\rho}\left[\cos(\rho-\pi\ell/2)
-\frac{(\ell+1)!}{(\ell-1)!}\frac{1}{2\rho}\sin(\rho-\pi\ell/2)\right]
+{\cal O}(\rho^{-3}) 
\label{nasymp}
\end{equation}
for a large $\rho$. This yields the suitable asymptotic form, 
\begin{equation}
f_{k,\ell}(r)\sim \sqrt{2}\sin(kr+\eta_\ell(k)-\pi\ell/2)+
\frac{(\ell+1)!}{(\ell-1)!}\frac{1}{\sqrt{2}kr}
\cos(kr+\eta_\ell(k)-\pi\ell/2),
\end{equation}
for $\ell\ge 1$. The case with $\ell=0$ is much simpler. 
Substituting the expression (\ref{exprfkell}) of $f_{k,\ell}$ into 
the left-hand side of (\ref{IkellR}), one has 
\begin{eqnarray}
I_{k,\ell}^R&=&\frac{d\eta_\ell(k)}{dk}+
R\cos^2\eta_\ell(k)\left[\left({\tilde j}'_\ell(kR)\right)^2-{\tilde j}_\ell(kR)
{\tilde j}''_\ell(kR)\right]\ret
&+&R\sin^2\eta_\ell(k)\left[\left({\tilde n}'_\ell(kR)\right)^2
-{\tilde n}_\ell(kr){\tilde n}''_\ell(kR)\right]\ret
&-&R\sin\eta_\ell(k)\cos\eta_\ell(k)
\left[2{\tilde j}'_\ell(kR){\tilde n}'_\ell(kR)-{\tilde j}_\ell(kR){\tilde n}''_\ell(kR)
-{\tilde j}''_\ell(kr){\tilde n}_\ell(kr)\right]\ret
&-&\frac{1}{k}\cos^2\eta_\ell(k){\tilde j}_\ell(kR){\tilde j}'_\ell(kR)
-\frac{1}{k}\sin^2\eta_\ell(k){\tilde n}_\ell(kR){\tilde n}'_\ell(kR)\ret
&+&\frac{1}{k}\sin\eta_\ell(k)\cos\eta_\ell(k)\left[{\tilde j}_\ell(kR){\tilde n}'_\ell(kR)
+{\tilde n}_\ell(kR){\tilde j}'_\ell(kR)\right],
\end{eqnarray}
where we have written ${\tilde j}_\ell(\rho)=\rho j_\ell(\rho)$ and 
${\tilde n}_\ell(\rho)=\rho n_\ell(\rho)$, and used the Wronskian, 
\begin{equation}
{\tilde j}_\ell(\rho){\tilde n}'_\ell(\rho)-{\tilde n}_\ell(\rho)
{\tilde j}'_\ell(\rho)=1. 
\end{equation}
Here $f'$ is the derivative of the function $f$.  
Taking the difference between the two quantities $I_{k,\ell}^R$ with and 
without the impurity potential $V(r)$, one obtains 
\begin{equation}
\int_0^R dr\left\{\left|f_{k,\ell}(r)\right|^2
-\left|f_{k,\ell}^{(0)}(r)\right|^2\right\}
=\frac{d\eta_\ell(k)}{dk}+g_\ell(k,R),
\label{deriveta}
\end{equation}
where 
\begin{eqnarray}
g_\ell(k,R)&:=&
-R\sin^2\eta_\ell(k)\left[\left({\tilde j}'_\ell(kR)\right)^2-{\tilde j}_\ell(kR)
{\tilde j}''_\ell(kR)\right]\ret
&+&R\sin^2\eta_\ell(k)\left[\left({\tilde n}'_\ell(kR)\right)^2
-{\tilde n}_\ell(kr){\tilde n}''_\ell(kR)\right]\ret
&-&R\sin\eta_\ell(k)\cos\eta_\ell(k)
\left[2{\tilde j}'_\ell(kR){\tilde n}'_\ell(kR)-{\tilde j}_\ell(kR){\tilde n}''_\ell(kR)
-{\tilde j}''_\ell(kr){\tilde n}_\ell(kr)\right]\ret
&+&\frac{1}{k}\sin^2\eta_\ell(k)\left[{\tilde j}_\ell(kR){\tilde j}'_\ell(kR)
-{\tilde n}_\ell(kR){\tilde n}'_\ell(kR)\right]\ret
&+&\frac{1}{k}\sin\eta_\ell(k)\cos\eta_\ell(k)\left[{\tilde j}_\ell(kR){\tilde n}'_\ell(kR)
+{\tilde n}_\ell(kR){\tilde j}'_\ell(kR)\right].
\end{eqnarray}
As is well known, the function $f_{k,\ell}(r)$ is continuous\footnote{
The continuity of the function $f_{k,\ell}$ can be proved by using 
the method of integral equation. See, for example, Chap.~12 of 
the book~\cite{Newton} of Newton.} with respect to 
$r\ge 0$ and to $k>0$, and the function $g_\ell(k,R)$ is also continuous 
with respect to $k>0$. Thus the existence of the derivative of $\eta_\ell(k)$ 
in the right-hand side of (\ref{deriveta}) is justified for $k>0$. 
Clearly this implies that the function $f_{k,\ell}$ is also differentiable 
with respect to $k>0$ for a large $r$ from the expression (\ref{exprfkell}). 
We decompose the function $g_\ell(k,R)$ into two parts as 
\begin{equation}
g_\ell(k,R)={\tilde g}_\ell(k,R)-\frac{1}{k}\sin\eta_\ell(k)
\cos(2kR+\eta_\ell(k)-\pi\ell).
\end{equation}
Note that the phase shift $\eta_\ell(k)$ satisfies\footnote{See, for example, 
Chap.~12 of the book \cite{Newton}.} 
\begin{equation}
|\sin\eta_\ell(k)|\le{\rm Const.}\cases{|k| & for $\ell=0$;\cr
                             |k|^{2\ell-1} & for $\ell\ge 1$,\cr}
\label{phaseshiftbound}
\end{equation}
and that the Bessel functions behave as 
\begin{equation}
{\tilde j}_\ell(\rho)\sim a\rho^{\ell+1}+b\rho^{\ell+3}
\end{equation}
and 
\begin{equation}
{\tilde n}_\ell(\rho)\sim c\rho^{-\ell}+d\rho^{-\ell+2}
\end{equation}
for a small $\rho$, where $a,b,c,d$ are the real constants. 
Using these properties, one can prove that the function ${\tilde g}_\ell(k,R)$ 
is bounded for any $k\in[0,k_{\rm F}]$. 
Further, by using the asymptotics (\ref{jasymp}) and (\ref{nasymp}) 
for a large $\rho$, one has 
\begin{equation}
{\tilde g}_\ell(k,R)\rightarrow 0 \quad \mbox{as}\quad R\rightarrow \infty
\quad \mbox{for a fixed }\ k\ne 0.
\end{equation}
{From} these observations, the dominated convergence theorem and (\ref{deriveta}), 
we obtain  
\begin{eqnarray}
& &\lim_{R\rightarrow\infty}
\int_0^R dr\int_0^{k_{\rm F}}dk\left\{\left|f_{k,\ell}(r)\right|^2
-\left|f_{k,\ell}^{(0)}(r)\right|^2\right\}\ret
&=&\eta_\ell(k_{\rm F})-\eta_\ell(0)
-\lim_{R\rightarrow\infty}\int_0^{k_{\rm F}}dk\frac{\sin\eta_\ell(k)}{k}
\cos(2kR+\eta_\ell(k)-\pi\ell),
\label{asymto} 
\end{eqnarray}
where the phase shift with $k=0$ is defined as 
$\eta_\ell(0):=\lim_{k\downarrow 0}\eta_\ell(k)$. 
The oscillating integral in the right-hand side 
vanishes in the limit $R\uparrow\infty$ 
by the Riemann-Lebesgue theorem and the boundedness of $\sin\eta_\ell(k)/k$ which 
holds because of the bound (\ref{phaseshiftbound}).  
Combining this result with (\ref{ZbetaR})--(\ref{ukellm}), one has 
\begin{equation}
\lim_{R\rightarrow\infty}Z_{\beta,R}=\sum_{\rm bound\ states}e^{-\beta\ell(\ell+1)}
+\frac{1}{\pi}\sum_{\ell=0}^\infty(2\ell+1)e^{-\beta\ell(\ell+1)}
\left[\eta_\ell(k_{\rm F})-\eta_\ell(0)\right].
\label{limZbetaR}
\end{equation}
We define 
the sum about $\ell$ in the right-hand side of (\ref{FSR}) 
by the limit $\beta\downarrow 0$ of the second sum in the right-hand side of 
(\ref{limZbetaR}). Then the desired result (\ref{FSR}) is obtained 
in the limit $\beta\downarrow 0$. 

%%%%%%%%%%%%%%%%%%%%%%%%%%%%%%%%%%%%%%%%%%%%%
\bigskip\bigskip\bigskip

\noindent
{\bf Acknowledgements:} We would like to thank Jacques Friedel for helpful 
comments. TK also thanks Daisuke Ida, Shu Nakamura, 
Hal Tasaki and Kenji Yajima for useful conversations. 
\bigskip\bigskip

%%%%%%%%%%%%%%%%%%%%%%%%%%%%%%%%%%%%%%%%%%%%%%%%%%%%%%%

\end{document}